\documentclass[11pt]{article}
\usepackage{booktabs}
\usepackage{PRIMEarxiv}
\usepackage{graphicx}
\usepackage{url}
\usepackage{microtype}
\usepackage{tabularx}
\usepackage{amsmath}
\usepackage[colorlinks=true,linkcolor=blue,citecolor=blue,urlcolor=blue]{hyperref}

\title{Beyond F1: Evaluating Coverage and Failure Recovery in AI Model Security Scanners}
\author{
Qianlong Lan, Vinothini Pandurangan, Anuj Kaul, Indranil Sanyal \\
 eBay Inc. \\
}
\date{}
\begin{document}
\maketitle

\begin{abstract}
Static scanners are increasingly used to identify executable or otherwise unsafe content in machine-learning artifacts, yet conventional evaluation metrics characterize only cases where a scanner yields a usable security judgment. We evaluate ModelScan, ModelAudit, and Fickling using a controlled, artifact-backed benchmark on a synthetic corpus of 170 Pickle and PyTorch focused artifacts across 145 specimen families, 135 of which have binary security ground truth and 10 of which are intentionally malformed without labels. We explicitly distinguish non-N/A coverage, analysis completion, definitive security decisions, non-security findings, and unsupported outcomes. On labeled families, ModelAudit produced definitive security decisions for all 135 families (100\%), Fickling for 110 (81.5\%), and ModelScan for 67 (49.6\%). Conditional on making a definitive judgment, ModelScan achieved 100\% precision, recall, and F1. Fickling identified no unique true-positive families beyond those found by the combination of ModelAudit and ModelScan. Furthermore, for the 48 malicious families where ModelScan failed to complete its analysis, both ModelAudit and Fickling generated detections consistent with ground truth. These findings underscore the need to separate judgment accuracy from judgment availability, as well as incremental detection coverage from tool-level redundancy.
\end{abstract}


\section{Introduction}

Model distribution creates a critical security boundary: artifacts may contain serialized objects, custom operators, embedded payloads, or malformed structures that exploit parser vulnerabilities. Python explicitly warns that Pickle data from untrusted sources can execute arbitrary code during unpickling~\cite{python-pickle}. PyTorch serialization inherits this vulnerability surface and documents restricted loading mechanisms such as weights\_only ~\cite{pytorch-serialization}. Model sharing platforms therefore deploy artifact scanning as an essential layer of supply-chain defense~\cite{hf-pickle-scanning,hf-protectai}.

We conduct a controlled, offline benchmark of three open-source model-security scanners: ModelScan, ModelAudit, and Fickling. Rather than reducing scanner output to a simple binary safe/unsafe decision, we explicitly track non-N/A coverage, successful completion, semantic security findings, incomplete analysis, and execution errors. Our analysis further evaluates family-level overlap, renamed-artifact robustness, operational latency, and the capacity of heterogeneous scanners to recover a defensible security decision when a primary scanner fails. This design addresses a central systems-security question: whether combining heterogeneous open-source scanners expands security coverage, improves decision availability under scanner failure, or merely introduces redundant confirmation.

Our evaluation is motivated by the insight that scanner quality encompasses two distinct dimensions: whether a scanner is correct when it yields a security judgment, and whether it can produce such a judgment in the first place. Standard conditional detection metrics capture the former but mask the latter. Consequently, we treat incomplete, unsupported, and structurally relevant outcomes as explicit, measurable operational states.

This paper makes three contributions:
\begin{enumerate}
    \item We define a family-aware evaluation methodology that separates
    observed non-N/A coverage, analysis completion, definitive security decisions,
    non-security findings, and execution failures. We report both corpus-wide
    and labeled-family decision coverage alongside conventional conditional
    detection metrics.
    \item We apply this methodology to a frozen synthetic benchmark of
    ModelScan, ModelAudit, and Fickling comprising 170 Pickle/PyTorch-focused
    artifacts organized into 145 specimen families.
    \item We distinguish incremental detection coverage from failure-mode
    redundancy through family-level scanner-overlap and recovery analysis, and
    additionally characterize false-positive causes, execution latency, and
    filename/extension robustness.
\end{enumerate}

\begin{figure*}[ht]
    \centering
    \includegraphics[width=\textwidth]{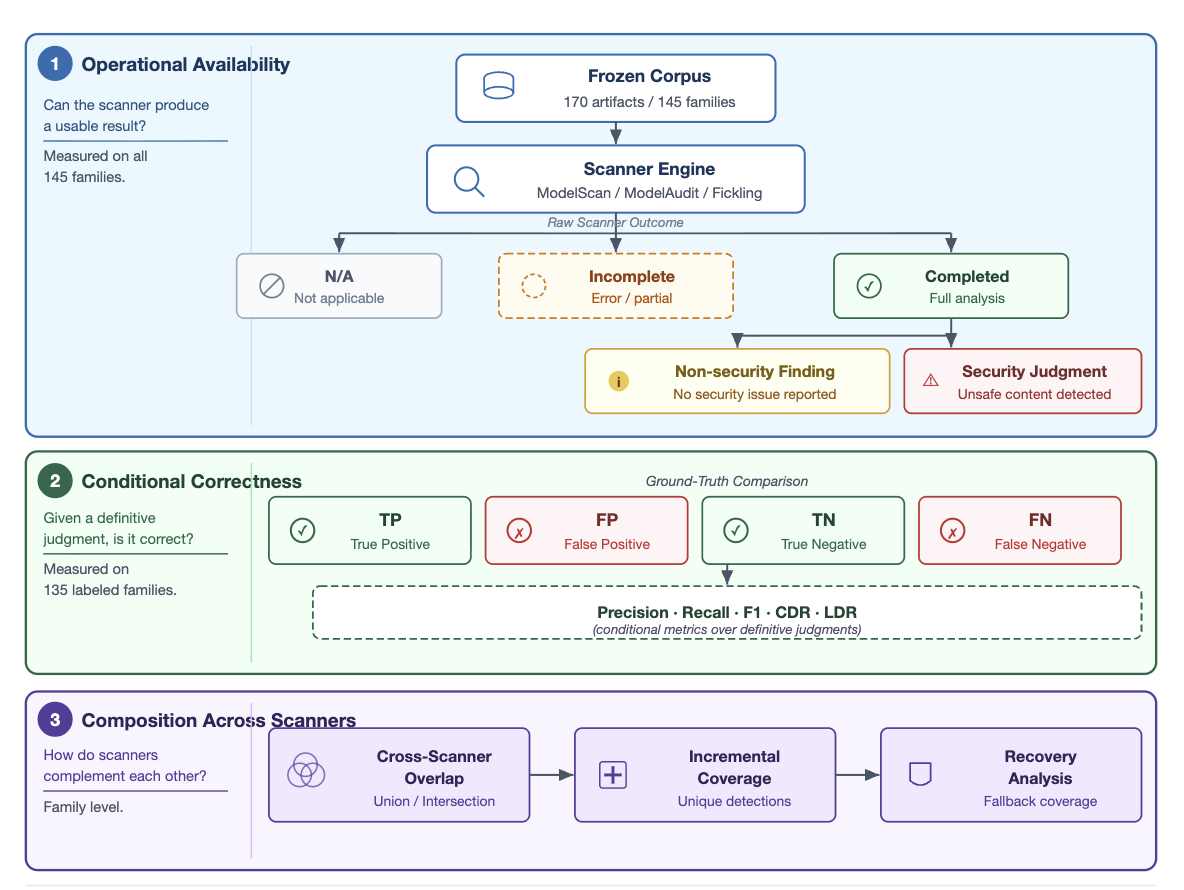}
    \caption{Benchmark evaluation model separating judgment availability,
    conditional correctness, and cross-scanner composition.}
    \label{fig:evaluation-model}
\end{figure*}

Figure~\ref{fig:evaluation-model} summarizes the benchmark's three analytical layers: whether a scanner produces a usable judgment, whether that judgment is correct relative to specimen-level ground truth, and whether multiple scanners provide incremental coverage or recovery.

\section{Research Questions}

We investigate the following research questions:
\begin{enumerate}
    \item \textbf{RQ1: Observed coverage, completion, and decision availability.}
    How do the scanners differ in observed non-N/A coverage, analysis completion,
    and their ability to produce definitive security judgments?
    \item \textbf{RQ2: Conditional accuracy versus practical coverage.}
    To what extent do conditional precision, recall, and F1 characterize
    scanner effectiveness when incomplete and unsupported outcomes are
    reported separately?
    \item \textbf{RQ3: Incremental value of specialized Pickle analysis.}
    Does Fickling contribute unique family-level true-positive or
    false-positive coverage beyond ModelAudit plus ModelScan?
    \item \textbf{RQ4: Operational recovery.}
    When one scanner produces an incomplete result, can heterogeneous scanners
    provide a ground-truth-consistent security decision?
    \item \textbf{RQ5: Operational characteristics.}
    How do the scanners differ in execution latency, and are their decisions
    stable under the evaluated filename and extension transformations?
\end{enumerate}

\section{Background and System Context}

Our evaluation is motivated by a production model-security pipeline where scanner workers operate asynchronously. To ensure reproducibility, we isolate the scanning layer and exclude deployment infrastructure and commercial platform integrations from the primary experiments. The benchmark thus characterizes isolated scanner behavior rather than an end-to-end model ingestion service or downstream execution sandbox.

\section{Methodology}

\subsection{Corpus and Experimental Conditions}

The primary experiment uses a frozen corpus of 170 artifacts organized into
145 specimen families. Families are the primary unit of security-effectiveness
analysis. Five families contain five artifacts each and five families contain
two artifacts each; the remaining 135 families contain one artifact.
Intentional renamed variants are grouped with their underlying specimen family
so that byte-identical copies are not counted as independent security
observations. Each artifact is inventoried and SHA-256 hashed before scanning.

The corpus contains 70 malicious families, 65 benign families, and 10
malformed families whose security ground truth is intentionally left unknown.
The malicious families comprise 60 unsafe-deserialization cases and 10
polyglot serialization cases. The benign families comprise 50 benign controls
and 15 benign checkpoints. The 10 unknown families are malformed/truncated
Pickle artifacts. All specimens are synthetic deterministic constructions or
transformed variants thereof.

The benchmark is deliberately focused on Pickle/PyTorch serialization. It
contains 130 Pickle artifacts, 30 PyTorch checkpoint artifacts, and 10 Pickle
container artifacts. Formats outside this domain, including standalone
SafeTensors, ONNX, and GGUF, are excluded.

The evaluated scanners are ModelScan~\cite{modelscan}, ModelAudit
~\cite{modelaudit}, and Fickling~\cite{fickling}. Versions are fixed at
ModelScan 0.8.6, ModelAudit 0.2.52, and Fickling 0.1.12. The controlled
benchmark executes offline with telemetry disabled, two workers, and
scanner-specific timeouts of 60, 10, and 5 seconds, respectively. The final
controlled run comprises 510 scanner/artifact executions with no missing,
duplicate, or unexpected execution combinations.

\subsection{Security Condition, Ground Truth, and Semantic Adjudication}

For this benchmark, a binary security-positive family is one whose frozen
construction label is \emph{malicious}: either an unsafe-deserialization
specimen designed to exercise executable reconstruction behavior or a
polyglot-serialization specimen constructed to carry the evaluated malicious
behavior. A binary security-negative family is a frozen benign control or
benign checkpoint. Malformed/truncated families have intentionally unknown
security ground truth and are excluded from TP/FP/TN/FN calculations.

Ground truth is assigned at the specimen-family level from the frozen corpus
manifest rather than inferred from scanner outputs. Transformed variants
retain the ground-truth label of their parent family. Scanner observations are
adjudicated against these frozen labels using the reviewed ledger bundled with
the benchmark.

A scanner warning is classified as a security finding when it asserts the
security condition evaluated by the benchmark. Structural, licensing, or
otherwise unrelated observations are classified as
\texttt{non\_security\_finding}. An analysis that does not establish a
definitive security judgment is classified as \texttt{scan\_incomplete}.
All 510 scanner observations have confirmed review status in the frozen
adjudication ledger.

We distinguish \emph{specimen maliciousness} from
\emph{deserialization capability}. Specimen maliciousness denotes whether the
frozen artifact implements the malicious behavior encoded by the benchmark
construction. Deserialization capability denotes whether an artifact contains
a primitive that a conservative security policy may regard as potentially
executable or unsafe. The confusion matrices in this paper evaluate specimen
maliciousness rather than deserialization capability.

This distinction is particularly important for benign custom serialization.
A scanner may conservatively flag a construct because it can participate in
executable deserialization even though the frozen specimen is intentionally
benign. Such a warning is security-relevant, but under the specimen-level
ground truth used here it is counted as a false positive. We therefore
characterize these false positives separately rather than interpreting them
as arbitrary scanner mistakes.

\subsection{Outcome Semantics}

We distinguish raw scanner observations from benchmark security interpretation.
Scanner-specific adapters parse process exit status, stdout, stderr, and
reported findings. A nonzero process exit status is not automatically
interpreted as scanner failure: valid detections emitted on a nonzero exit are
retained.

Each scanner/family observation is assigned to one of:
\begin{itemize}
    \item \texttt{detected}: completed analysis with an applicable security finding;
    \item \texttt{clean}: completed analysis with no applicable security finding;
    \item \texttt{not\_applicable}: no claimed coverage for the artifact format;
    \item \texttt{scan\_incomplete}: no definitive judgment was established;
    \item \texttt{scanner\_error}: scanner execution or output parsing failed;
    \item \texttt{non\_security\_finding}: an observation did not correspond to
    the benchmark security condition.
\end{itemize}

\subsection{Metrics}

Let $N=145$ denote all specimen families and $N_L=135$ the families with
binary security ground truth.

\paragraph{Observed non-N/A coverage.}
We operationalize scanner format applicability through the scanner's explicit
disposition: a family contributes to observed non-N/A coverage when the
scanner does not return \texttt{not\_applicable}. We report $A_s/N$. This
metric captures the absence of an explicit unsupported-format disposition; it
must not be interpreted as demonstrated successful format support, because a
non-N/A analysis may still be incomplete or erroneous.

\paragraph{Analysis completion.}
Among families with non-N/A dispositions, completion means that analysis terminates
with a scanner result rather than an incomplete or scanner-error state.
Completion may yield either a definitive security judgment or a structural /
non-security observation.

\paragraph{Decision coverage.}
Let
\[
D_s = TP_s + FP_s + TN_s + FN_s.
\]
We report two complementary rates:
\[
CDR_s = \frac{D_s}{N},
\qquad
LDR_s = \frac{D_s}{N_L}.
\]
We call $CDR_s$ the \emph{corpus-wide definitive security-judgment rate} and
$LDR_s$ the \emph{labeled-family decision rate}. The latter removes the ten
families whose benchmark security ground truth is intentionally unknown.
Neither metric converts incomplete, not-applicable, scanner-error, or
non-security outcomes into false negatives.

\paragraph{Conditional detection metrics.}
For definitive judgments,
\[
P_s=\frac{TP_s}{TP_s+FP_s}, \qquad
R_s=\frac{TP_s}{TP_s+FN_s},
\]
\[
F1_s=2\frac{P_sR_s}{P_s+R_s}.
\]
We additionally report specificity and false-positive rate over benign
families when the relevant denominator is defined:
\[
\mathrm{Specificity}_s=\frac{TN_s}{TN_s+FP_s}, \qquad
FPR_s=\frac{FP_s}{TN_s+FP_s}.
\]

\paragraph{Operational recovery.}
Scanner $s_j$ recovers scanner $s_i$ on family $f$ when $s_i$ returns an
incomplete result and $s_j$ produces a definitive judgment on the same family
that agrees with benchmark ground truth.

\subsection{Family-Level and Artifact-Level Analysis}

Security-effectiveness metrics use family-level analysis to avoid
pseudoreplication from intentional variants. Artifact-level results are
retained for routing and rename experiments, where each variant is an
intentional observation.

\subsection{Latency and Rename Evaluation}

Latency is measured for 170 executions per scanner. Timing includes scanner
process invocation and artifact analysis. The rename experiment contains 15
scanner-level comparisons spanning five malicious Pickle families and three
scanners. Underlying bytes remain unchanged while filename/extension
declarations vary, including \texttt{.txt}, \texttt{.py}, \texttt{.bin}, and
an omitted extension.

\section{Results}

\subsection{Observed Coverage, Completion, and Decision Availability}

Table~\ref{tab:operational} separates observed non-N/A coverage, completion, and
security-decision availability. ModelAudit has non-N/A dispositions for all 145
families and completes all 145 analyses, but 10 completed observations concern
the unknown malformed subset and are therefore not binary security judgments.
ModelScan has non-N/A dispositions for all families but leaves 78 family-level
analyses incomplete. Fickling has non-N/A dispositions for 120 families and
completes all 120, with 10 completed observations belonging to the unknown
malformed subset.

\begin{table}[htbp]
\centering
\caption{Family-level operational coverage. Completion is shown both
conditional on a non-N/A disposition and relative to the full corpus.
``Definitive'' counts judgments eligible for TP/FP/TN/FN classification.}
\label{tab:operational}
\setlength{\tabcolsep}{4pt}
\begin{tabular}{lrrrrr}
\toprule
Scanner & Non-N/A & Completed/non-N/A & Completed/all & Incomplete & Definitive \\
\midrule
ModelAudit & 145/145 & 145/145 & 145/145 & 0  & 135 \\
ModelScan  & 145/145 & 67/145  & 67/145  & 78 & 67 \\
Fickling   & 120/145 & 120/120 & 120/145 & 0  & 110 \\
\bottomrule
\end{tabular}
\end{table}

\begin{table}[htbp]
\centering
\caption{Frozen family-level security results. Conditional metrics are
calculated only over definitive security judgments.}
\label{tab:family-results}
\setlength{\tabcolsep}{2.8pt}
\begin{tabular}{lrrrrrrrrr}
\toprule
Scanner & TP & FP & TN & FN & Non-sec. & Inc. & N/A & CDR & LDR \\
\midrule
ModelAudit & 70 & 60 & 5 & 0 & 10 & 0 & 0 & 93.1\% & 100\% \\
ModelScan  & 22 & 0 & 45 & 0 & 0 & 78 & 0 & 46.2\% & 49.6\% \\
Fickling   & 60 & 50 & 0 & 0 & 10 & 0 & 25 & 75.9\% & 81.5\% \\
\bottomrule
\end{tabular}
\end{table}

\begin{figure}[htbp]
\centering
\includegraphics[width=0.76\linewidth]{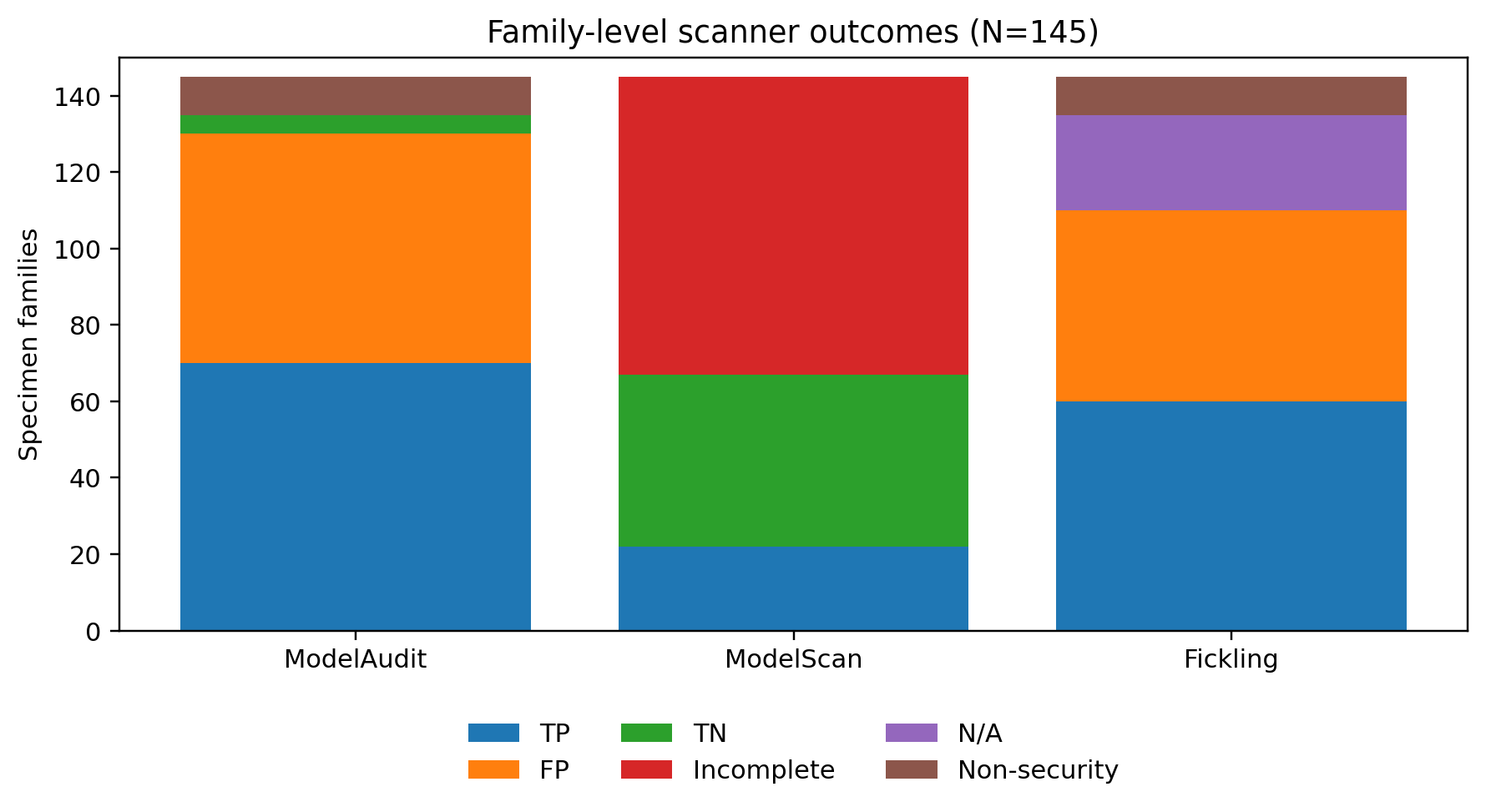}
\caption{Family-level scanner outcomes. The ten malformed families have
unknown security ground truth and appear as non-security or incomplete
outcomes rather than TP/FP/TN/FN.}
\label{fig:outcomes}
\end{figure}

Figure~\ref{fig:outcomes} visualizes the outcome composition. It makes clear
that conditional accuracy and judgment availability are distinct properties.

\subsection{Conditional Accuracy Can Obscure Incompleteness}

ModelScan achieved 100\% conditional precision, recall, and F1 on its 67
definitive security judgments, but its labeled-family decision rate is only
67/135 (49.6\%). Thus, the perfect conditional score does not imply that
ModelScan produces a security judgment for the full labeled corpus.
Equivalently, conditional F1 describes the correctness of ModelScan's returned
judgments but provides no information about the remaining 68/135 labeled
families for which no definitive security judgment is available.

The corpus-wide rate is lower still, 67/145 (46.2\%), because it retains the
ten unknown malformed families in the operational denominator. Reporting both
rates avoids treating the benchmark's absence of binary ground truth as if it
were a scanner failure.

\subsection{Benign-Specimen Warning Characterization}

The largest alert-quality difference concerns benign custom serialization.
ModelAudit produces 60 false-positive families: all 50 benign controls and 10
of the 15 benign checkpoints. It correctly classifies the remaining five
benign checkpoint families. Its specificity over the 65 benign families is
5/65 (7.7\%), corresponding to an FPR of 60/65 (92.3\%).

Fickling produces 50 false-positive families, all drawn from the 50 benign
control families. The 15 benign checkpoint families are not applicable to
Fickling. Among the 50 benign families for which Fickling produces definitive
judgments, its specificity is therefore 0/50 and its conditional FPR is
100\%. ModelScan produces no false positives among its definitive benign
judgments: it classifies 45 benign families as true negatives and leaves 20
benign families incomplete.

The adjudication ledger assigns the same reviewed cause to the ModelAudit and
Fickling false positives: each scanner emitted a security warning for a frozen
benign custom serialization. This result should therefore be interpreted in
light of scanner semantics. Conservative identification of potentially
executable deserialization constructs can be security-relevant even when the
specific frozen specimen is benign. Our FP label measures disagreement with
specimen-level ground truth, not whether the warning is operationally useless.

\begin{table}[htbp]
\centering
\caption{Benign-family warning dispositions under specimen-level ground truth.}
\label{tab:fp}
\begin{tabular}{lrrrr}
\toprule
Scanner & Benign FP & Benign TN & Benign incomplete/N/A & FPR$^\dagger$ \\
\midrule
ModelAudit & 60 & 5 & 0 & 92.3\% \\
ModelScan  & 0 & 45 & 20 & 0\% \\
Fickling   & 50 & 0 & 15 & 100\% \\
\bottomrule
\end{tabular}

\smallskip
\begin{minipage}{0.92\linewidth}
\footnotesize
$^\dagger$ FPR is conditional on definitive benign judgments. For ModelScan
and Fickling, incomplete or not-applicable benign families are excluded from
the FPR denominator.
\end{minipage}
\end{table}

\subsection{Fickling Adds Redundancy Without Incremental Family Coverage}

Fickling produces 60 true-positive and 50 false-positive families. Every
Fickling TP is already detected by ModelAudit or ModelScan, and every Fickling
FP is already present in their false-positive union. Fickling therefore adds
zero unique TP and zero unique FP families beyond the ModelAudit--ModelScan
union in this corpus.

\subsection{Cross-Scanner Recovery of ModelScan Incompleteness}

ModelScan has 78 incomplete family outcomes: 48 malicious, 20 benign, and 10
unknown malformed families. The malicious subset is the security-critical
recovery case. ModelAudit and Fickling each produce true-positive detections
for all 48/48 malicious families on which ModelScan is incomplete. ModelAudit
alone therefore achieves the same malicious-family recovery as the
three-scanner configuration.

For the 20 benign ModelScan-incomplete families, ModelAudit produces warnings
that are false positives under specimen-level ground truth; Fickling also
warns on the applicable benign controls while benign checkpoint families may
be not applicable. Consequently, the 48/48 malicious recovery result should
not be generalized into a claim that heterogeneous scanners correctly recover
all incomplete decisions.

\subsection{Union-Ensemble Coverage}

For the union policy used here, a family is flagged when \emph{any} scanner in
the configured set flags it. This is a detection union, not majority voting or
consensus.

\begin{table}[htbp]
\centering
\caption{Family-level union coverage. Adding scanners to ModelAudit does not
change the observed malicious-family coverage or benign false-positive union.}
\label{tab:ensemble}
\begin{tabular}{lrr}
\toprule
Scanner set & Malicious detected & Benign falsely flagged \\
\midrule
ModelAudit & 70/70 & 60/65 \\
ModelAudit $\cup$ ModelScan & 70/70 & 60/65 \\
All three scanners & 70/70 & 60/65 \\
\bottomrule
\end{tabular}
\end{table}

Because ModelAudit already detects all 70 malicious families, no union
configuration containing ModelAudit can improve true-positive family coverage
on this frozen corpus. The ensemble experiment therefore primarily measures
redundancy and false-positive inheritance rather than additional TP coverage.
Under this union policy, adding ModelScan or Fickling to ModelAudit does not
change the observed TP set and does not reduce ModelAudit's false-positive set.

\subsection{Rename Robustness}

All 15 evaluated scanner-level rename comparisons preserve the scanner
decision. The evaluated transformations include misleading declarations such
as \texttt{.txt}, \texttt{.py}, \texttt{.bin}, and an omitted extension.
This is a routing/filename-sensitivity result only and does not establish
robustness to content obfuscation, archive rewriting, or serialization changes.

\subsection{Operational Latency}

\begin{table}[htbp]
\centering
\caption{Scanner invocation latency under the frozen benchmark configuration.}
\label{tab:latency}
\begin{tabular}{lrr}
\toprule
Scanner & Median & P95 \\
\midrule
Fickling & 64.90 ms & 80.98 ms \\
ModelAudit & 963.21 ms & 2.08 s \\
ModelScan & 14.30 s & 16.26 s \\
\bottomrule
\end{tabular}
\end{table}

These measurements include process invocation and artifact analysis. The
frozen run records Python 3.12.13, two workers, offline execution, telemetry
disabled, and scanner timeouts of 5, 10, and 60 seconds for Fickling,
ModelAudit, and ModelScan, respectively. CPU, memory, and storage metadata were
not retained. We therefore treat these values as within-run operational
observations rather than portable scanner-performance benchmarks.

\subsection{Research-Question Answers}
The preceding analyses answer the five research questions from complementary
operational and security perspectives. The scanners differ substantially in
decision availability despite strong conditional detection performance on
the judgments they return. Additional scanners do not expand the observed
family-level detection set once ModelAudit is included, but they can provide
redundant detections when another scanner is incomplete. The evaluated rename
transformations preserve scanner decisions, while the latency measurements
show substantial differences in operational cost under the frozen
configuration. Table~\ref{tab:research-questions} summarizes these findings
and their corresponding research questions.

\begin{table}[htbp]
\centering
\caption{Answers from the frozen benchmark.}
\label{tab:research-questions}
\begin{tabularx}{\linewidth}{p{0.34\linewidth}X}
\toprule
Question & Frozen answer \\
\midrule
Non-N/A/completion/coverage? &
ModelAudit has non-N/A dispositions for and completes 145/145 families; ModelScan has
non-N/A dispositions for 145/145 but completes 67/145; Fickling has non-N/A
 dispositions for 120/145 and completes all 120 non-N/A analyses. Labeled-family decision
rates are 100\%, 49.6\%, and 81.5\%. \\
Is conditional F1 sufficient? &
No; ModelScan has 100\% conditional F1 but a 49.6\% labeled-family decision rate. \\
Does Fickling add unique TP/FP coverage? &
No; it adds zero unique TP and zero unique FP families beyond ModelAudit plus ModelScan. \\
Can scanners recover incomplete analysis? &
For malicious ModelScan-incomplete families, yes: ModelAudit and Fickling each detect 48/48. \\
Are rename decisions stable? &
Yes for the evaluated subset; all 15 scanner-level comparisons preserve decisions. \\
\bottomrule
\end{tabularx}
\end{table}

\section{Discussion}
Our primary empirical finding is that conditional detection quality and judgment availability reflect fundamentally different operational properties. While ModelScan achieves zero false positives and false negatives across its 67 definitive judgments, it yields a definitive security decision for only 49.6\% of labeled families. Conversely, ModelAudit achieves 100\% labeled-family decision coverage and detects all 70 malicious families, though its 92.3\% benign-family FPR under specimen-level ground truth introduces a substantial operational alert burden.

These results also highlight the importance of separating benchmark ground-truth semantics from scanner design objectives. Both ModelAudit and Fickling conservatively flag benign custom serializations because they possess executable-deserialization capabilities. Although these warnings are categorized as false positives relative to frozen specimen labels, they reveal an inherent policy tradeoff between blocking structurally hazardous constructs and allowing known-benign custom serializations. Similar precision-coverage trade-offs are well established across static security analysis literature~\cite{aloraini2019,lipp2022}.

Furthermore, the overlap analysis differentiates detection complementarity from failure-mode redundancy. Although Fickling does not expand family-level true-positive coverage beyond the combination of ModelAudit and ModelScan, it successfully detects all 48 malicious families within the ModelScan-incomplete subset. Integrating supplementary scanners can therefore provide fallback decision availability under tool failures without necessarily expanding the union's primary detection coverage.

Deployment policies must consequently look beyond raw scanner counts to consider marginal detection yields, failure-recovery capacity, false-positive overhead, format applicability, and operational cost. Finally, our findings suggest that evaluation frameworks should explicitly separate conservative capability-flagging metrics from specimen-level maliciousness benchmarks when their underlying semantics diverge.

\section{Limitations and Safety}

This benchmark operates under a strictly controlled scope and does not represent a population-level estimate of model-scanner performance across arbitrary production environments. The dataset of 145 families and 170 artifacts focuses specifically on Pickle and PyTorch serialization formats.

All corpus artifacts are synthetic, deterministic constructions or transformed variants. While this design ensures experimental reproducibility and strict ground-truth control, it cannot substitute for real-world malware, adaptive adversaries, or previously unseen zero-day exploits. Furthermore, the class distribution within the benchmark is structured for diagnostic evaluation rather than sampled from deployment prevalence; conditional precision figures should therefore not be interpreted as expected production positive predictive values.

The 10 malformed families intentionally lack binary security labels. By reporting both corpus-wide and labeled-family decision coverage, our methodology prevents these incomplete or unparseable artifacts from being silently misclassified as scanner execution failures or valid ground-truth cases.

False positives are evaluated against specimen-level security labels. For tools designed to conservatively flag potentially executable serialization primitives, certain benchmark false positives may still provide actionable security warnings. Although our ledger analysis explicitly distinguishes structural capabilities from specimen maliciousness, quantifying deployment-specific operational utility requires dedicated policy studies.

In addition, scanner evaluations represent fixed version snapshots. The rename robustness experiment is limited to 15 scanner-level comparisons across five families, and Fickling's zero-incremental coverage finding remains specific to this corpus composition. Static serialization scanning also does not guarantee the absence of neural weight-space backdoors or ensure runtime security during downstream model execution. Finally, because hardware telemetry was not recorded during the frozen run, latency metrics should be interpreted strictly as internal operational observations rather than cross-platform benchmarks.

\section{Reproducibility}

The complete benchmark artifact is publicly available at
\url{https://github.com/lanqianlong/AI-Model-Security-Scanners}. The repository
contains the frozen benchmark bundle and experimental materials required to
inspect and reproduce the results reported in this paper.

The frozen bundle records the complete experimental provenance, including the
corpus manifest, artifact SHA-256 checksums, scanner software versions, all
510 execution records, raw scanner outputs, reviewed adjudication entries,
the family-level scanner matrix, recovery analysis, rename evaluations, and
mechanically computed metrics. The controlled execution matrix contains zero
duplicate execution IDs, zero missing or unexpected scanner--artifact
combinations, and valid references to the retained raw outputs. The benchmark
was executed offline with telemetry disabled.

All 510 adjudication records have confirmed review status. To preserve
experimental integrity, the corpus manifest and adjudication ledger are frozen
and cryptographically hashed in the benchmark metadata. Primary results and
summary metrics are derived mechanically from these frozen artifacts rather
than entered manually.

The public repository also provides the corpus-generation tooling, scanner
adapters, benchmark configuration, and metric-generation scripts used by the
evaluation. Together with the frozen manifests and recorded scanner versions,
these materials support independent inspection and reproduction of the
reported benchmark results. Because scanner dependencies and upstream software
may evolve, the frozen versions and integrity metadata should be used when
attempting to reproduce the results reported here.

\section{Related Work}

\subsection{Serialization and Model Supply-Chain Security}

The security hazards of dynamic deserialization are well documented. Python's official documentation explicitly warns that unpickling untrusted data can trigger arbitrary code execution~\cite{python-pickle}. PyTorch inherits these security risks due to its reliance on Pickle, prompting official documentation for restricted loading mechanisms such as weights\_only ~\cite{pytorch-serialization}. Model repositories operationalize these concerns at scale: platforms like Hugging Face integrate static Pickle scanning and third-party security tools as primary components of their supply-chain defense infrastructure~\cite{hf-pickle-scanning,hf-protectai}.

\subsection{Model Security Scanners and Safe Loading}

Recent open-source tools address model artifact security through static inspection and safe loading. ModelScan performs static AST analysis to identify unsafe deserialization calls~\cite{modelscan}, Fickling specializes in decompiling and analyzing Pickle bytecode sequences~\cite{fickling}, and ModelAudit inspects model file structures for embedded security risks~\cite{modelaudit}. In contrast to pure detection scanners, PickleBall enforces safe loading policies during model deserialization~\cite{pickleball}. Our evaluation complements these efforts by systematically characterizing operational scanner behavior—specifically analyzing non-N/A coverage, completion rates, false-positive semantics, decision coverage, tool overlap, recovery capacity, and execution latency.

\subsection{Security-Scanner Benchmarking}

Static security analyzer evaluations consistently demonstrate that detection count alone cannot summarize tool efficacy. Aloraini et al.\ empirically evaluate false-positive warnings in SAST tools, highlighting the operational burden of high alert rates~\cite{aloraini2019}. Lipp et al.\ benchmark C static analyzers against real-world vulnerabilities, showing that combining tools can reduce false negatives while simultaneously increasing total flagged code~\cite{lipp2022}. We apply a similar operational perspective to machine learning artifact scanners, introducing explicit operational states for unsupported formats and analysis incomplete outcomes alongside cross-scanner recovery mechanisms.

\subsection{Threats Beyond Serialization}

Executable serialization payloads represent only one vector within the broader model supply-chain threat landscape. Static serialization scanners cannot establish the absence of neural weight-space backdoors or data-poisoning artifacts. We treat these orthogonal threat vectors as out of scope for our benchmark rather than conflating them with deserialization scanner effectiveness.
\section{Conclusion}

This benchmark demonstrates that detection accuracy conditional on successful analysis is insufficient to evaluate model-artifact security scanners. Across 135 specimen families with binary security ground truth, ModelScan achieves 100\% conditional F1 but yields definitive security judgments for only 49.6\% (67/135) of families. ModelAudit produces judgments across all 135 labeled families and detects all 70 malicious families, but falsely flags 60 out of 65 benign families. Fickling provides judgments for 110 labeled families, contributing no unique true-positive detections beyond the combination of ModelAudit and ModelScan.

Furthermore, scanner ensembles exhibit two distinct operational behaviors. Although Fickling does not expand overall family-level detection coverage, it successfully detects all 48 malicious families where ModelScan analysis remains incomplete. Integrating supplementary scanners can therefore offer critical failure-mode redundancy without necessarily increasing unique detection yield.

These findings highlight the necessity of evaluating scanners along both judgment correctness and judgment availability while explicitly characterizing false-positive warning semantics. Future work will extend this framework to naturally occurring malicious artifacts, adversarial evasion techniques, broader serialization ecosystems, longitudinal software versions, and deployment-level policy designs.

\bibliographystyle{plain}
\bibliography{references}
\end{document}